\documentstyle[preprint,aps]{revtex}

\draft
\newcommand{\be}{\begin{equation}}
\newcommand{\ee}{\end{equation}}
\newcommand{\ben}{\begin{eqnarray}}
\newcommand{\een}{\end{eqnarray}}

\newcommand{\iii}{\'{\i}}

\newcommand{\nn}{\nonumber \\ }

\begin{document}
\draft
\title{The statistics of the entanglement
changes generated by the Hadamard-CNOT quantum circuit}
\author{J. Batle$^1$, M. Casas$^1$, A. Plastino$^{2,\,3}$,
 and A. R. Plastino$^{1,\,2,\,4}$}

\address {
$^1$Departament de F\iii sica, Universitat de les Illes Balears,
07071 Palma de Mallorca, Spain \\
  $^2$Argentina's National Research Council (CONICET) \\
$^3$Department of Physics, National University La Plata,
  C.C. 727, 1900 La Plata, Argentina \\
  $^4$Faculty of Astronomy and Geophysics, National University La Plata,
  C.C. 727, 1900 La Plata, Argentina
}

\date{\today}

\maketitle

\begin{abstract}
We consider the change of entanglement of formation $\Delta E$ produced by the
Hadamard-CNOT circuit  on a general (pure or mixed) state $\rho$ describing a
system of two qubits. We study numerically the probabilities of obtaining
different values of $\Delta E$, assuming that the initial state is randomly
distributed in the space of all states according to the product measure
recently introduced by Zyczkowski {\it et al.} [Phys. Rev. A {\bf 58} (1998)
883].

\vskip 5mm
 Pacs: 03.67.-a; 89.70.+c;
03.65.Bz
\end{abstract}
\vspace{.5cm}

\newpage
 Entanglement is one of the most fundamental phenomena of quantum mechanics
\cite{LPS98}. It is a physical resource, like energy, associated with the
peculiar non-classical correlations that are possible between separated quantum
systems. One needs entanglement so as  to implement quantum information
processes \cite{WC97,W98,BEZ00,AB01,PV98,M02,SH00,DSB02} such as quantum
cryptographic key distribution \cite{E91}, quantum teleportation
\cite{BBCJPW93}, superdense coding \cite{BW93}, and quantum computation
\cite{EJ96,BDMT98,galindo}. Production of entanglement is the elementary
prerequisite for any quantum computation. This basic task is accomplished by
unitary transformations $\hat U$  (quantum gates) representing quantum
evolution acting on the space state of multipartite systems. $\hat U$ should
describe nontrivial interactions among the degrees of freedom of its
subsystems.

   One of the  fundamental questions about quantum computation is then
   how to construct an adequate set of
quantum gates, and a nice answer can be given: any generic two-qubits gate
suffices for universal computation \cite{barenco1}.  One would then be
legitimately interested in ascertaining just how efficient distinct $\hat U$'s
are as entanglers. In this respect, much exciting work has recently been
performed  (see, for instance, \cite{ZZF00,Z01,KC01,DVCLP01,WZ02,CO00,EI00}).

A state of a composite quantum system is called ``entangled" if it can not be
represented as a mixture of factorizable pure states. Otherwise, the state is
called separable. The above definition is physically meaningful because
entangled states (unlike separable states) cannot be prepared locally by acting
on each subsystem individually \cite{We89,P93}.
  A physically motivated measure of entanglement is
 provided by the entanglement of formation $E[\rho]$  \cite{BDSW96}. This measure
 quantifies the resources needed to create a given entangled state $\rho$. That is,
 $E[\rho]$ is equal to the asymptotic limit (for large $n$) of the
 quotient $m/n$, where $m$ is the number of singlet states needed to create $n$
 copies of the state $\rho$ when the optimum procedure based on local
 operations is employed. The entanglement of formation for two-qubits
 systems is given by Wootters' expression  \cite{WO98},

\be
E[\rho] \, = \, h\left( \frac{1+\sqrt{1-C^2}}{2}\right), \ee

\noindent where

\be
h(x) \, = \, -x \log_2 x \, - \, (1-x)\log_2(1-x), \ee

\noindent and $C$ stands for the {\it concurrence}  of the two-qubits state
$\rho$. The concurrence is given by

\be
C \, = \, max(0,\lambda_1-\lambda_2-\lambda_3-\lambda_4), \ee

\noindent $\lambda_i, \,\,\, (i=1, \ldots 4)$ being the square roots, in
decreasing order, of the eigenvalues of the matrix $\rho \tilde \rho$, with

\be \label{rhotil} \tilde \rho \, = \, (\sigma_y \otimes \sigma_y) \rho^{*}
(\sigma_y \otimes \sigma_y). \ee

\noindent The above expression has to be evaluated by recourse to the matrix
elements of $\rho$ computed  with respect to the product basis.
%%%%%%%%%%%%%%%%%%%%%%%%%%%%%%%%%%%%%

In the present effort we will concern ourselves with one
particular quantum circuit: the Hadamard-CNOT circuit, that
combines two gates: a single-qubit one (Hadamard's) with a
two-qubits gate (CNOT).  Quantum logic gates are unitary evolution
operators $\hat U$ that act on the states of a certain number of
qubits. If the number of such qubits is $m$, the quantum gate is
represented by a $2^m$x$2^m$ matrix in the unitary group $U(2^m)$.
These gates are reversible: one can reverse the action, thereby
recovering an initial quantum state from a final one. We shall
work here with $m=2$.
 The simplest nontrivial 2-qubits
operation is the quantum controlled-NOT, or CNOT (equivalently, the exclusive
OR, or XOR). Its classical counterpart is a reversible logic gate operating on
two bits: $e_1$, the control bit,  and $e_2$, the target bit. If $e_1=1$, the
value of $e_2$ is negated. Otherwise, it is left untouched. The quantum CNOT
gate $C_{12}$ (the first subscript denotes the control bit, the second the
target one)  plays a paramount role in both experimental and theoretical
efforts that revolve around the quantum computer concept. In a given
orthonormal basis $\{\vert 0 \rangle,\,\vert 1 \rangle\}$, and if we denote
addition modulo 2 by the symbol $\oplus$, we have \cite{barenco}

\be \label{uno} \vert e_1 \rangle\, \vert e_2 \rangle \rightarrow
C_{12}\rightarrow \vert e_1 \rangle\, \vert e_1 \oplus e_2 \rangle.   \ee
%&
In conjunction with simple single-qubit operations, the CNOT gate constitutes a
set of gates out of which {\it any quantum gate may be built} \cite{barenco1}.
In other words, single qubit and CNOT gates are universal for quantum
computation \cite{barenco1}.

As stated, the CNOT gate  operates on quantum states of two qubits and is
represented by a 4x4-matrix. This matrix has a diagonal block form. The upper
diagonal block is just the unit 2x2 matrix. The lower diagonal 2x2 block is the
representation of the one-qubit NOT gate $U_{NOT}$, of  the form

\ben \label{block}  0\,\,\,\,\,1 \nn   1\,\,\,\,\,0
 \een

\noindent
 Note that, of course, $C_{12}^2=1$. This gate is
  able to transform factorizable pure states into entangled ones,
 i.e., \be \label{enta} C_{12}: [c_1 \vert 0 \rangle + c_2 \vert
1 \rangle] \vert 0 \rangle \leftrightarrow c_1 \vert 0 \rangle
\vert 0 \rangle + c_2 \vert 1 \rangle \vert 1 \rangle, \ee and
this transformation can be reversed by applying the same CNOT
operation once more \cite{barenco}.

The Hadamard transform $T_H$ ($T_H^2=1$) is given by

\be \label{Hada} T_H= \frac{1}{\sqrt{2}} [\sigma_1 + \sigma_3] ,\ee and acts on
the single qubit basis $\{ |0>,\,\, |1> \}$ in the following fashion

\ben \label{hadamad} T_H |0> &=& \frac{1}{\sqrt{2}}[|1> - |0>] \nn T_H |1> &=&
\frac{1}{\sqrt{2}}[|0> + |1>] .\een \noindent Consider now the two-qubits
uncorrelated basis $\{ |00>,\,  |01>,\,  |10>,\,  |11>
 \}.$ If we act with $T_H$ on the members of this basis we obtain
 \ben \label{basisun}
 && \frac{1}{\sqrt{2}}\,\,\big[|1> - |0>\big]\,\,\,\,\,\,|0> \nn
&& \frac{1}{\sqrt{2}}\,\,\big[|1> - |0>\big]\,\,\,\,\,\,|1> \nn
&&\frac{1}{\sqrt{2}}\,\,\big[|0> + |1>\big]\,\,\,\,\,\,|0>\nn
&&\frac{1}{\sqrt{2}}\,\,\big[|0> + |1>\big]\,\,\,\,\,\,|1>,  \een
\noindent so that the posterior action of the CNOT gate yields
\newpage \ben \label{basiscorr}
 &&\frac{1}{\sqrt{2}}\,[ |1>|1> - |0>|0> ] \nn
 &&\frac{1}{\sqrt{2}}\,[ |1>|0> - |0>|1> ] \nn
 &&\frac{1}{\sqrt{2}}\,[ |0>|0> + |1>|1> ] \nn
 &&\frac{1}{\sqrt{2}}\,[ |0>|1> + |1>|0> ],\een
i.e., save for an irrelevant overall phase factor in two of the kets, the
maximally correlated Bell basis $\vert \phi^{\pm} \rangle$, $\vert \psi^{\pm}
\rangle$. We see then that the $T_H$-CNOT combination transforms an
uncorrelated basis in the maximally correlated one.

Now, the  two-qubits systems with which  we are going to be concerned here are
the simplest quantum mechanical systems exhibiting the entanglement phenomenon
and play a fundamental role in quantum information theory. The concomitant
space ${\cal H}$ of {\it mixed states} is 15-dimensional and its properties are
not of a trivial character. While the entanglement of pure states can be
regarded as well understood, the entanglement of mixed states still has many
properties that deserve further investigation. The reason for this state of
affairs lies in the fact the quantum content of the associated correlations is
somewhat obscured by the classical correlations in a mixed state. A mixed state
 which does not violate any Bell inequality can nonetheless exhibit quantum
 mechanical correlations, as one can distill from it pure maximally entangled
 states that violate Bell inequalities \cite{PV98}.

There are still then ${\cal H}$-features, related to the phenomenon of
entanglement, that have not yet been characterized in full detail. One such
characterization problem will occupy us here. We shall perform a systematic
numerical survey of the action of  the $T_H$-CNOT circuit  on our
15-dimensional space in order to ascertain the manner in which $P(\Delta E)$ is
distributed in ${\cal H}$, with $P$ the probability of generating a change
$\Delta E$ associated to the action of this reversible quantum circuit. This
kind of exploratory work is in line with recent efforts towards the systematic
exploration of the space of arbitrary (pure or mixed) states of composite
quantum systems \cite{ZHS98,Z99,ZS01} in order to determine the typical
features exhibited by these states with regards to the phenomenon of quantum
entanglement \cite{ZHS98,Z99,ZS01,MJWK01,IH00,BCPP02a,BCPP02b}. It is important
to stress the fact that we are exploring a space in which the majority of
states are {\it mixed}. The exciting investigations reported in
\cite{ZZF00,Z01,KC01,DVCLP01,WZ02} address mainly pure states. We will try to
answer the question: given an initial degree of entanglement of formation $E$,
what is the probability $P(\Delta E)$ of encountering a change in entanglement
$\Delta E$ upon the action of this circuit?

Our answer will arise from a Monte Carlo exploration of ${\cal H}$. To do this
we need to define a proper measure on ${\cal H}$. The space of all (pure and
mixed) states $\rho$ of a quantum system described by an $N$-dimensional
Hilbert space can be regarded as a product space ${\cal S} = {\cal P} \times
\Delta$ \cite{ZHS98,Z99}. Here $\cal P$ stands for the family of all complete
sets of orthonormal projectors $\{ \hat P_i\}_{i=1}^N$, $\sum_i \hat P_i = I$
($I$ being the identity matrix). $\Delta$ is the set of all real $N$-uples
$\{\lambda_1, \ldots, \lambda_N \}$, with $0 \le \lambda_i \le 1$, and $\sum_i
\lambda_i = 1$. The general state in ${\cal S}$ is of the form $\rho =\sum_i
\lambda_i P_i$.  The Haar measure on the group of unitary matrices $U(N)$
induces a unique, uniform measure $\nu$ on the set ${\cal P}$
\cite{ZHS98,Z99,PZK98}. On the other hand, since the simplex $\Delta $ is a
subset of a $(N-1)$-dimensional hyperplane of ${\cal R}^N$, the standard
normalized Lebesgue measure ${\cal L}_{N-1}$ on ${\cal R}^{N-1}$ provides a
measure for $\Delta$. The aforementioned measures on $\cal P$ and
$\Delta$ lead to a measure $\mu $ on the set $\cal S$ of quantum states
\cite{ZHS98,Z99},

\be \label{memu}
 \mu = \nu \times {\cal L}_{N-1}.
 \ee

 We are going to consider the set of states of a two-qubits
 system. Consequently, our system will have $N=4$ and, for such an $N$, ${\cal S}\equiv {\cal
 H}$.
 All our present considerations are based on the assumption
 that the uniform distribution of states of a two-qubit system
 is the one determined by the measure (\ref{memu}). Thus, in our
 numerical computations we are going to randomly generate
 states of a two-qubits system according to the measure
 (\ref{memu}) and  study the entanglement evolution of these states
upon the action of our $T_H$-CNOT quantum circuit.

As a first step, we suggest that the reader take a look at Fig. 4a of Ref.
\cite{BCPP02a}. There one finds a plot of the probability $P(E)$ of finding
two-qubits states of ${\cal
 H}$  endowed with a given amount of entanglement $E$. In this
 graphs, the solid line corresponds to all states (pure and mixed), while the
 dashed curve depicts pure state behaviour only. We clearly see that our
 probabilities are of a quite different character when they refer to of pure
states than when they correspond to mixed ones. Most mixed states
have null entanglement, or a rather small amount of it (see the
enlightening discussion in \cite{ZHS98}). For pure states it is
more likely to encounter them endowed with an intermediate
(between null and total) amount of
 entanglement. It is then important to ascertain how much entanglement the  $T_H$-CNOT quantum
 circuit is able to generate on our 15-dimensional two-qubits space.

 We deal with pure states only in Fig. 1.
 %Fig. 1a plots the probability $P(E_F)$ of generating via the $T_H$-CNOT quantum
 %circuit  final state  with entanglement $E_F$ if the initial
 %entanglement is zero (solid line).
Fig. 1a plots the probability $P(\Delta E)$ of obtaining via the $T_H$-CNOT quantum
 circuit a final state with entanglement change $\Delta E=E_F-E_0$.
 In 1b we are concerned with the average value $\langle E_F
 \rangle$ pertaining to final states that result from the gate-operation on initial
 ones of a given (fixed) entanglement $E_0$ (solid line). The horizontal line
 is plotted for the sake of reference. It corresponds to the
 average entanglement of two-qubits pure states, equal to
 $1/(3\ln{2})$. The diagonal line $\langle E_F \rangle$ = $E_0$ is also shown (dashed line).
$E_F$ is a decreasing function of $E_0$ although the quantum circuit considered
increases the mean final entanglement approximately
up to 0.5 for states with $E_0$ lying in the interval $[0, 0.5]$.

The same analysis, but involving now all states (pure and mixed),
 is summarized in  Fig. 2. The graph 2a is the counterpart of 1a, while 2b is that of 2a.
 The dashed line of 2b, given for the sake of visual reference, if
 just the line $\langle E_F
 \rangle=E_0$.
 The two Figs. allow one to
 appreciate the fact that it is quite unlikely that we may generate,
via the $T_H$-CNOT quantum circuit,
 a significant amount of entanglement if the initial state is
 separable. In Fig 2 we see that  the
mean final entanglement
 $\langle E_F \rangle$ rises rapidly near the origin, from zero, with $E_0$ .
 The rate of entanglement-growth decreases steadily with $E_0$ and the interval
in which $\langle E_F \rangle$ is greater than $E_0$ is significantly smaller
that the one corresponding to pure states (Fig. 1b).
 %For pure states, though, the behavior is in this respect is quite
 %different.

The $P(\Delta E)$ vs. $\Delta E$ plots exhibit a  nitid peak at
$\Delta E=0$. The peak is enormously exaggerated if mixed states
enter the picture (2a). Thus, if the initial state has null
entanglement, our survey indicates that the most probable
circumstance is that the circuit will leave its entanglement
unchanged.

We performed a systematic survey, in the space of all two-qubits
states, concerning the entanglement changes associated with the
action of the $T_H$-CNOT circuit. We found that the probability
distribution of entanglement changes obtained when the circuit
acts on pure states is quite different from the distribution
obtained when the circuit acts on general mixed states. The
probability of entangling mixed states turns out to be rather
small. On average, the $T_H$-CNOT transformation is more
efficient, as entangler, when acting upon states with small
initial entanglement, specially in the case of pure states.

\acknowledgments
This work was partially supported by the DGES grant PB98-0124 (Spain),
and by CONICET (Argentine Agency).

\newpage

\noindent {\bf FIGURE CAPTIONS}

 \vskip 0.5cm

\noindent Fig. 1-a) $P(\Delta E)$ vs. $\Delta E$ for pure states.
The change of entanglement $\Delta E$ arises as a result of the
action of the $T_H$-CNOT quantum circuit. b) Probability of
obtaining, via the $T_H$-CNOT transformation, a final state with
entanglement $E_F$, when the initial state is endowed with a given
entanglement $E_0$ (solid line). The horizontal line depicts the
mean entanglement of all pure states. The diagonal (dashed line)
is drawn for visual reference. \vskip 0.5cm

\noindent Fig. 2 The same as in Fig. 1 for all states (pure and mixed).
%
%\vskip 0.5cm
%
%\noindent Fig. 3- a) $P(\Delta E)$ vs. $\Delta E$ for pure states.
%  The change of entanglement  $\Delta E$ arises as a result of
%  the acting of a CNOT gate (solid line) and a $\pi/4$-one (dashed curve). The
%  dotted-dashed curve reflects the  entanglement change  $\Delta E$ between
% two randomly chosen pure states. b) Corresponding results for all states. The
%solid line corresponds to the CNOT transformation and the dashed one to the
%$\pi/4$ gate.
%
%
%\vskip 0.5cm
%
%\noindent Fig. 4-a) Mean initial entanglement $\langle E_0 \rangle$ vs $\Delta
%E$ for pure states. b) For all states. c) $\langle E_0^2 \rangle$- $\langle E_0
%\rangle^2$ vs. $\Delta E$ for pure states. d) For all states. The solid line
%refers to the CNOT transformation and the dashed one to the $\pi/4$ gate.
%
%\vskip 0.5cm \noindent Fig. 5. Effects of the action of the CNOT gate (left)
%and of the $\pi/4$-gate (right). We plot $P(E_F)$ vs. $E_F$ for fixed initial
%entanglement $E_0= 0.1,\,0.2,\,0.3,$ and $0.4$, respectively.
%
%\vskip 5mm \noindent Fig. 6 a) Mean initial entanglement $\langle E_0 \rangle$
%vs $\Delta E$ via the CNOT gate for states with a given value of the
%participation rate $R= 1.4$ and $R=2.2$. b) Its associated fluctuation $\langle
%E_0^2 \rangle$- $\langle E_0 \rangle^2$ vs. $\Delta E$.

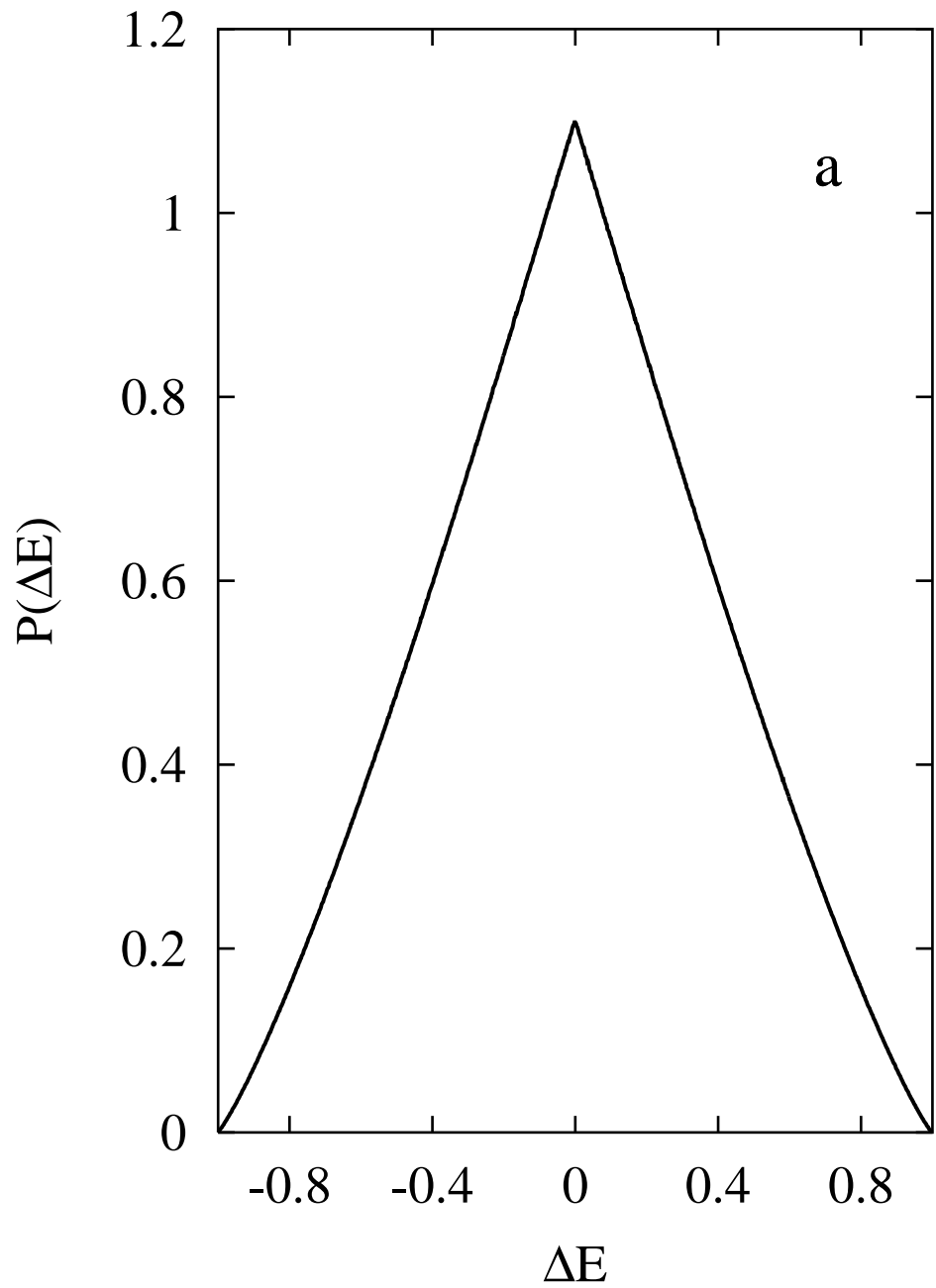
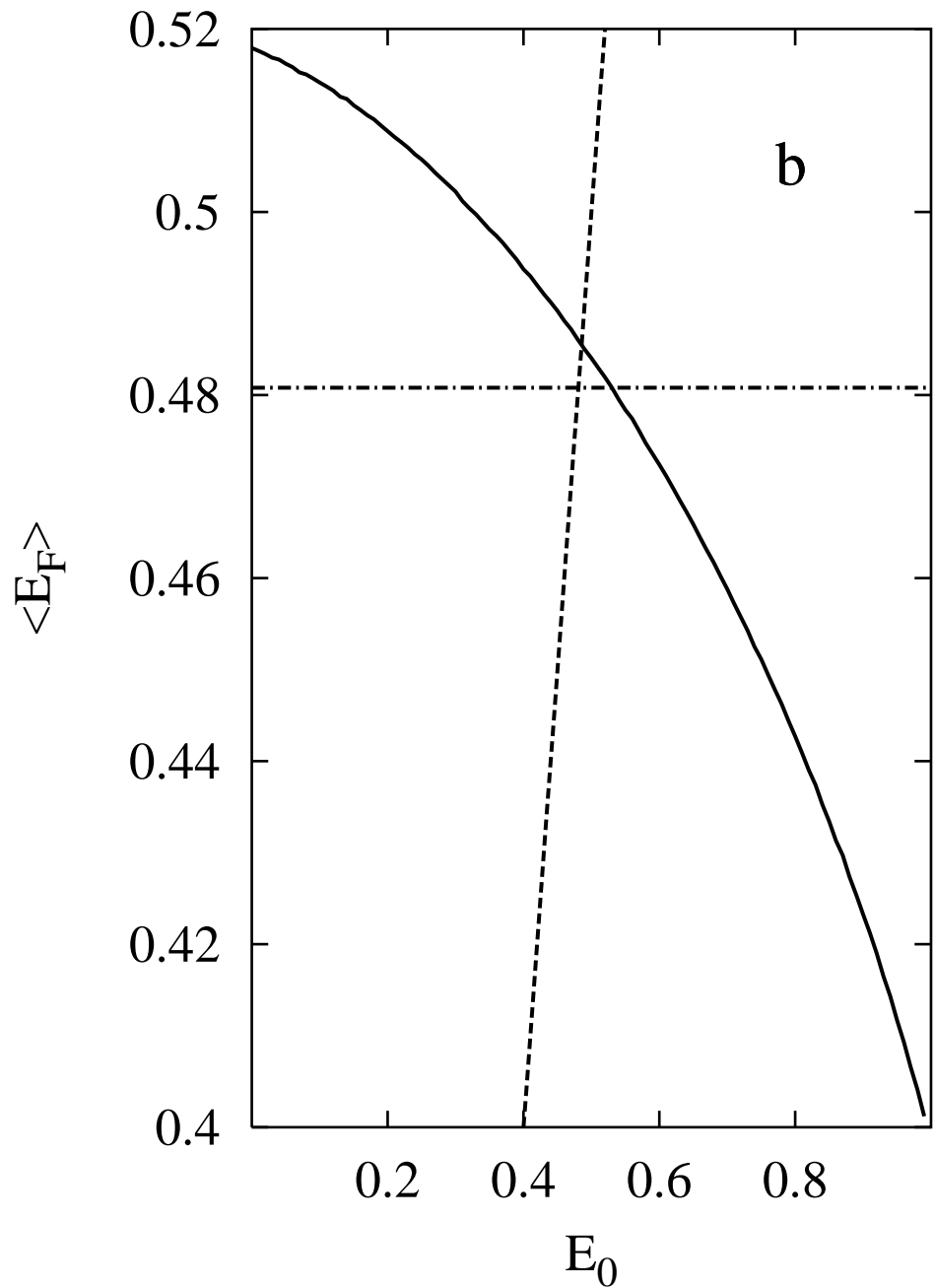
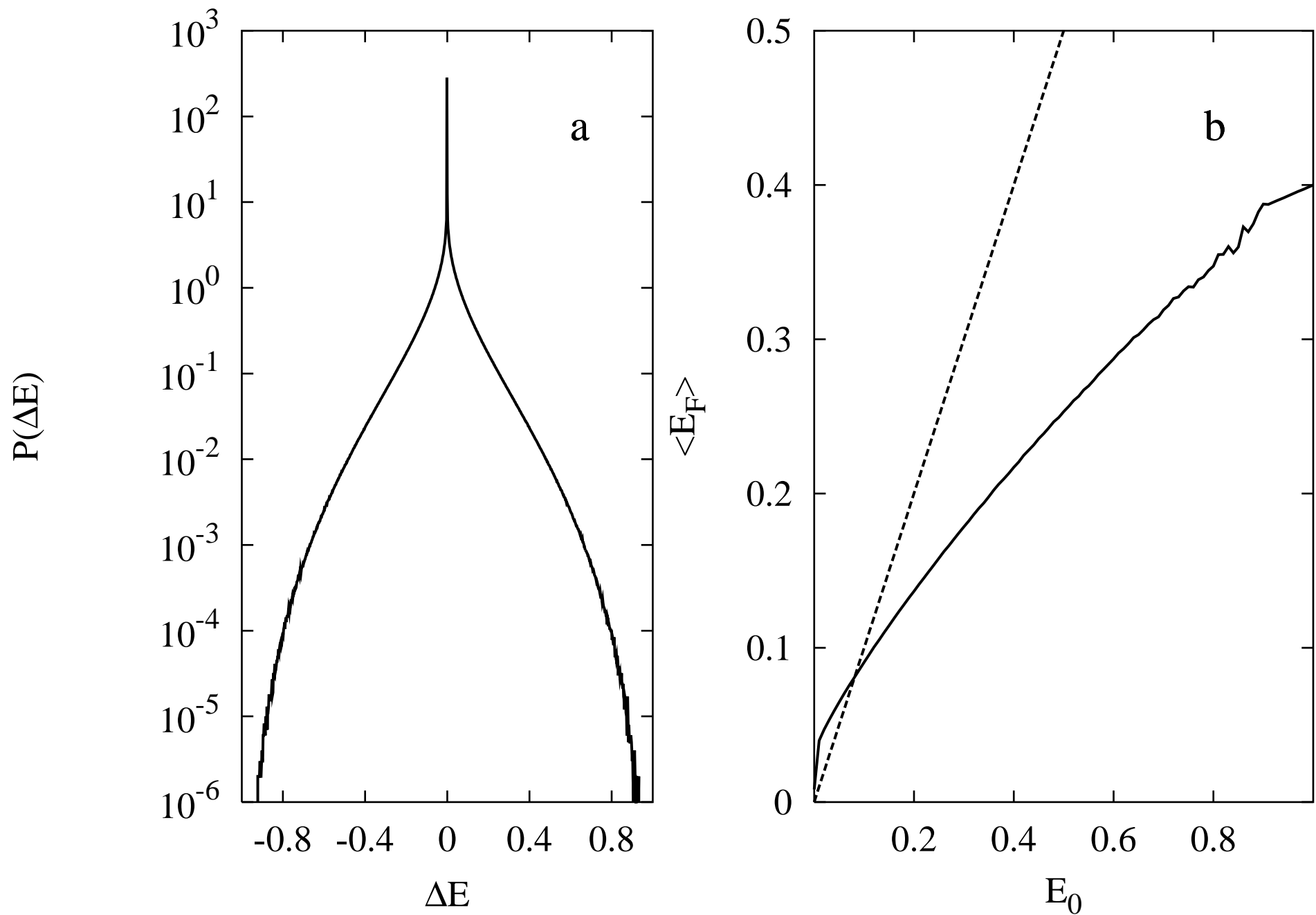
\end{document}